\documentclass{aa}
\usepackage{graphics}
\begin{document}

%%%%% AUTHORS - PLACE YOUR OWN MACROS HERE %%%%%

%\newcommand{\beq}{\begin{equation}}
%\newcommand{\eeq}{\end{equation}}

\def\bi{\bibitem}
%***************************** REFERENCES *****************************
%
% This template is useful for referencing articles in journals.
\newcommand{\journ}[5]{
                      {#1},       % authors
                      {#2}.       % year
                      {#3},   % journal
                      {#4},   % volume
                      {#5}.   % page
                      }
%
% And this one is useful if the article is submitted or in press.
\newcommand{\inpress}[4]{
                  {#1},        % authors
                  {#2},        % year
                  {#3\/},      % journal
                  {#4}.        % something like submitted, or in press.
                        }
%
% And this one is useful if the article is in preparation.
\newcommand{\inprep}[3]{
                  {#1},  % authors
                  {#2}.  % year
                  {#3}.  % something like "in preparation", or ...
                       }
%
% This template is useful for referencing books.
\newcommand{\book}[4]{
                   {#1},        % authors
                   {#2}.        % year
                   {\it #3\/}, % title of book
                   {#4}.        % place of publication/publisher
                   }
%
% This template is useful for referencing articles in proceedings.
\newcommand{\proceed}[6]{
                   {#1},        % authors
                   {#2}.        % year
                in {\it #3\/}, % title of book
              eds. {#4},        % editors
                   {#5},        % publisher
                p. {#6}.        % page
                        }
%
% This is used for proceedings not yet pressed.  
\newcommand{\proceedinpress}[5]{
                   {#1},        % authors
                   {#2},        % year
                in {\it  #3\/}, % title of book
              eds. {#4},        % editors
                   {#5}.        % comment, as in press
                               }
%
% This template is useful for referencing thesis or dissertations.
\newcommand{\thesis}[3]{
                   {#1},        % author
                   {#2},        % year
                   {\it  #3\/}, % type (Master, PhD) and institution
                       }

%%%%%%%%%%%%%%%%%%%END%%%%%%%%%%%%%%%%%%%%%%%%%%%%%

\thesaurus{11.13.1,11.19.4}
        
\title{Morphologies and ages of star cluster pairs and multiplets in the Small Magellanic Cloud} 
 
\author{M.R. de Oliveira\inst{1}
\and {C.M. Dutra}\inst{1}
\and { E. Bica}\inst{1}
\and {H. Dottori}\inst{1}}

\institute{Instituto de Fisica-UFRGS, CP 15051, CEP 91501-970 POA - RS, Brazil}

\offprints{marcio@if.ufrgs.br}
\date{Received 2 May 2000 / Accepted}

\titlerunning{Morphologies and Ages of Star Cluster Pairs and Multiplets in the SMC}

\authorrunning{M.R. de Oliveira  et al.}

\maketitle

\begin{abstract}
An isophotal atlas of 75 star cluster pairs and multiplets in the Small
Magellanic Cloud is presented, comprising 176 objects. They are concentrated in
the SMC main body. The isophotal contours were made from Digitized Sky
Survey$^{*}$ images and showed relevant structural features possibly related to
interactions in about 25\% of the sample. Previous N-body simulations indicate
that such shapes could be due to tidal tails, bridges or common envelopes.  The
diameter ratio between the members of a pair is preferentially in the range $1 -
2$, with a peak at $1$.  The projected separation is in the range $\approx$ $3 -
22$ pc with a pronounced peak at $\approx$ $13$ pc. For 91 objects
it was possible to derive ages from Colour-Magnitude Diagrams using the OGLE-II
photometric survey. The cluster multiplets in general occur in OB stellar
associations and/or HII region complexes.  This indicates a common origin and
suggests that multiplets coalesce into pairs or single clusters in a short time
scale. Pairs in the SMC appear to be mostly coeval and consequently
captures are a rare phenomenon. We find evidence that star cluster pairs and
multiplets may have had an important role in the dynamical history of clusters
presently seen as large single objects.

\keywords{Magellanic Clouds -- Star Clusters}

\end{abstract}

\thanks { $^*$The images in
this study are based on photographic data obtained using the UK Schmidt
Telescope, which was operated by the Royal Observatory Edinburgh, with funding
from the UK Science and Engineering Research Council, until 1988 June, and
thereafter by the Anglo-Australian Observatory.  Original plate material is
copyright by the Royal Observatory Edinburgh and the Anglo-Australian
Observatory. The plates were processed into the present compressed digital form
with their permission.  The Digitized Sky Survey was produced at the Space
Telescope Science Institute under US Government grant NAG W-2166.}

\section{Introduction}

Star cluster pairs are common objects in the Magellanic Clouds and it is
important to understand their formation and evolution processes.  A list of 30 cluster pairs
in the SMC was first presented by Hatzidimitriou \& Bhatia (1990). Bica \& Schmitt
(1995, hereafter BS95) revised previous data on SMC extended objects (star
clusters, associations and emission nebulae) and identified new ones using Sky
Survey ESO/SERC R and J films. They presented a list of 40 pairs and 2 triple
star clusters.  Pietrzy\'nski \& Udalski (1999a) reported 23 pairs and 4
triplets derived from Pietrzy\'nski et al.'s (1998) star cluster catalogue
in the OGLE survey area. Bica \& Dutra (2000) updated BS95's catalogue
considering the entries in Pietrzy\'nski et al's catalogue. Bica \& Dutra (2000)
indicated 75 pairs and multiplets comprising 176 individual objects.

In recent years there has been growing evidence of interacting star clusters in
 the Magellanic Clouds, especially in the LMC. Bhatia \& Hatzidimitriou (1988)
 concluded that more than 50\% of LMC pairs must be physical systems. Bhatia \&
 McGillivray (1988) found evidence that NGC2214 is a merging binary star
 cluster, based on the presence of a flattened core and an extended halo. Indeed
 Sagar et al. (1991) detected two turnoff points revealing the presence of
 two interacting clusters. Bica et al. (1992) studied cluster pairs and
 multiplets in the LMC bar by means of integrated colours and found systems which
 resulted coeval and some with  age differences. Vallenari et
 al (1998) confirmed such scenarios by means of colour-magnitude
 diagrams. They also presented isophotal contours indicating physical
 interaction. Several other studies have found binarity evidence in LMC cluster
 pairs (e.g. Kontizas  et al. 1993, Dieball \& Grebel 2000a, 2000b).

Bhatia  et al. (1991) presented a photographic atlas of binary star cluster
candidates in the LMC. For the SMC no morphological atlas is available and
isophotal maps are required to test possible physical interactions.  Indeed
comparisons of isophotes with isopleth maps of N-body simulations proved to be a
useful tool (Rodrigues et al. 1994, de Oliveira et al. 1998,
hereafter ODB98), since the simulations produce features such as bridges, common
envelopes and extensions. The observational importance of tidal tails as
interaction signatures was also indicated by Leon et al. (1999).

In this work we provide isophotal maps for SMC pairs and multiplets to study the
following properties of these candidate physical systems: (i) angular
distribution; (ii)  projected centre-to-centre separation of members; (iii) isophotal
structures using the Digitized Sky Survey$^*$ (hereafter DSS); (iv) ages
derived by means of isochrone fitting, when possible. We discuss candidate  physical systems
  and infer a scenario for their formation and evolution.

In Sect. 2 we gather the objects
providing coordinates, sizes, centre-to-centre separations and other details for
the SMC pairs and multiplets. In Sect. 3 we provide the isophotal contour
atlas together with classifications of interaction features whenever present. A
preliminary version of the present isophotal SMC atlas together with one  for the
LMC  was given in de Oliveira (1996).  In Sect. 4 we derive cluster ages by
means of Colour-Magnitude diagrams extracted from the OGLE-II BVI photometric
survey (Udalski et al. 1998).  In Sect. 5 we discuss the properties of
the systems and the possible scenarios for their origin and evolution. Finally,
concluding remarks are given in Sect. 6.

\section{SMC pairs and Multiplets}

Table~\ref{tab:catalog} shows data for the 176 SMC objects which form 75 star
 cluster pairs and multiplets (Bica \& Dutra 2000), considering that the pair BS63/B67
 in the latter study is possibly a triplet with NGC294 (ODB98). By columns: (1)
 object cross-identification in the different catalogues: N (Henize 1956), K
 (Kron 1956), L (Lindsay 1958), H (Hodge 1960), SL (Shapley \& Lindsay 1963), B
 (Br\"uck 1976), DEM (Davies et al. 1976), ESO (Lauberts 1982), H86-
 (Hodge 1986), MA (Meyssonier \& Azzopardi 1993), BS (Bica \& Schmitt 1995),
 OGLE (Pietrzy\'nski  et al.  1998). Note that some are embedded objects,
 named after the corresponding HII region; (2) and (3) right ascension and
 declination for the epoch J2000, respectively; (4) object type following BS95:
 C for star cluster, NC for small HII regions with embedded star clusters, CN
 for clusters which show some traces of emission, A for associations, CA and/or
 AC for star clusters of low density and objects with intermediate properties,
 AN for associations which show some traces of emission and NA for HII regions
 with embedded associations; (5) and (6) major and minor sizes, respectively;
 (7) position angle of major axis ($0^{\circ}$=N, $90^{\circ}$=E); (8)
 centre-to-centre angular separation (1 arcsec = 0.28 pc assuming an absolute
 distance modulus (m-M) = 18.9 for the SMC, Westerlund 1990). For triplets and
 multiplets we measured the separation between the two main members, as
 indicated in the corresponding object line in the table; (9) remarks: $m6, m5,
 m4, mT$ and $mP$ indicate groups with six, five and four members, triplets and
 pairs, respectively. A running number identifies each candidate
 system. Abbreviations `br*' indicates that a bright star is present; `att'
 means attached to. A hierarchical indication is given for objects embedded in
 or superimposed on larger ones: `in' suggests a possible physical connection
 while `sup' suggests a projection. There are 56 pairs, 15 triplets, 2
 quadruplets, 1 quintuplet and 1 sextuplet. The quintuplet is located in the
 star forming region NGC395 with dimensions 25 pc x 18 pc which in turn is
 embedded in the HII complex SMC-DEM126 with 66 pc x 35 pc according to angular
 sizes in BS95. The sextuplet is mostly contained in the OB association H-A35
 (Hodge 1985) with dimensions 57 pc x 47 pc. One quadruplet is located in H-A60
 with 44 pc x 24 pc. Note that dimensions (columns 5 and 6) and position angle (column 7) are from
 Bica \& Dutra (2000) and were measured on plates. The dimensions are in general larger than the
isophotal sizes in the subsequent analyses. 

\begin{table*}
\caption[]{SMC pairs and multiplets}
\begin{scriptsize}
\renewcommand{\tabcolsep}{0.9mm}
\begin{tabular}{l r r r r r r r l}
\hline
\hline
Name & RA(2000) & Dec(2000)& T & Dmax & Dmin & PA & Separ. & Comments \cr
     & h : m : s &$^{o}$ : ' : ''&  & '' & '' & $^{o}$& '' & \cr 
(1)  & (2)   & (3)       &(4) & (5) & (6) & (7) & (8) & (9) \cr 
\hline
 
  BS3 			  & 0:30:01 & -73:20:01 & AC & 33 & 33 & - & 31 & mT-1 \cr 
  H86-2 		  & 0:30:04 & -73:20:35 & AC & 39 & 36 & 80 & 31 & mT-1 \cr 
  BS4 			  & 0:30:07 & -73:20:58 & C & 33 & 27 & 70 & & mT-1 \cr
 
  BS7                     &  0:32:43 & -73:37:59 &   C & 33 & 33 &    -  &  59 & mP-1      \cr
  BS8                     &  0:32:56 & -73:38:33 &  AC & 36 & 36 &    -  &       &  mP-1     \cr
                                                     
  H86-22                  &  0:34:53 & -73:02:08 &   C & 18 & 18 &    -  & 42 & mP-2 \cr
  BS9                     &  0:35:02 & -73:02:14 &  AC & 45 & 39 &   100 &       & mP-2 \cr
 
  HW9                     &  0:36:25 & -73:00:05 &   C & 45 & 45 &    -  &    52   &mP-3  \cr
  HW10                    &  0:36:31 & -72:59:13 &   C & 57 & 57 &    -  &       & mP-3 \cr 
 
  B9                      & 0:37:13  & -72:57:53 &   C & 27 & 27 &    -  &   68    &mP-4  in H-A1 \cr
  H86-41                  & 0:37:29  & -72:57:48 &   C & 33 & 24 &    50 &       &mP-4   \cr 

  H86-38                  & 0:37:24  & -73:01:50 &   A & 72 & 72 &    -  &   46    &mP-5   \cr
  BS10                    & 0:37:34  & -73:01:30 &   C & 39 & 39 &    -  &       &mP-5   \cr 
                                                                                                             
  BS14,SMC\_OGLE165        & 0:39:12  & -73:14:46 &   C & 33 & 33 &    -  &  44  &mP-6   \cr
  SMC\_OGLE5               & 0:39:22  & -73:15:28 &  CA & 51 & 45 &    90 &  &mP-6   \cr   
                                                         
  HW12A                   & 0:39:26  & -73:22:59 &   C & 33 & 27 &    10 & 36 & mP-7   \cr
  H86-54                  & 0:39:35  & -73:22:58 &   C & 33 & 33 &    -  &  &mP-7   \cr 

  B15                     & 0:39:42  & -72:58:39 &  CA & 36 & 27 &   100 & 30 &mP-8   \cr
  H86-57                  & 0:39:47  & -72:58:56 &  CA & 30 & 30 &    -  &  &mP-8   \cr  
                                                                                                         
  BS13                    & 0:40:08  & -72:45:30 &  CA & 54 & 45 &    40 &  48 &mP-9   \cr
  BS248                   & 0:40:15  & -72:46:02 &  AC & 39 & 30 &    70 &  &mP-9   \cr  
                                                     
  NGC220,K18,L22,ESO29SC3 &  0:40:31 & -73:24:10 &   C & 72 & 72 &     - &  88 & mT-2,in H-A3  \& SMC\_OGLE8 \cr
  NGC222,K19,L24,ESO29SC4 &  0:40:44 & -73:23:00 &   C & 72 & 72 &     - &  88 & mT-2,in H-A3  \& SMC\_OGLE9  \cr
  B23,SMC\_OGLE170         &  0:40:55 & -73:24:07 &   C & 36 & 30 &    40 &  &mT-2,in H-A3  \cr
 
  B19                     &  0:40:44 & -73:03:42 &   C & 27 & 21 &   170 &  32 &mT-3 \cr
  B20                     &  0:40:49 & -73:04:09 &   C & 18 & 18 &     - &  32 & mT-3 \cr
  H86-62,SMC\_OGLE10       &  0:40:48 & -73:05:17 &   C & 36 & 36 &    -  &  &mT-3  \cr  
                                                    
  NGC231,K20,L25,ESO29SC5 &  0:41:06 & -73:21:07 &   C & 108 & 108 &    -  &  36 &mP-10,in H-A3  \& SMC\_OGLE11  \cr 
  BS15                    &  0:41:21 & -73:20:31 &   A & 138 & 138 &    -  &  &mP-10,in H-A3 \cr
                                            
  B21,SMC\_OGLE171         &  0:41:15 & -72:49:55 &   C & 21 & 21 &    -  & 34  &mP-11   \cr
  B22                     &  0:41:21 & -72:49:32 &   C & 21 & 21 &    -  &  &mP-11   \cr     
                                                 
  B29                     &  0:42:11 & -73:43:51 &  CA & 30 & 21 &   100 & 46 &mP-12   \cr 
  HW16,SMC\_OGLE13         &  0:42:22 & -73:44:03 &  CN & 36 & 36 &    -  &  &mP-12,in SMC-DEM7 \cr
                                        
  NGC241,K22w,L29W,       &  0:43:33 & -73:26:25 &   C & 57 & 57 &    -  & 23 &mP-13 \& ESO29SC6w,SMC\_OGLE17 \cr
  NGC242,K22e,L29e,       &  0:43:38 & -73:26:37 &   C & 45 & 45 &    -  &  &mP-13 \& ESO29SC6e,BH1,SMC\_OGLE18 \cr
                                   
  B31,SMC\_OGLE19,SMC\_OGLE175 &  0:43:38 & -72:57:31  &  C  & 30 & 24 & 150 &  &mT-4 \cr
  BS20,SMC\_OGLE20            &  0:43:38 & -72:58:48  &  C  & 27 & 27 &  -  & 30 &mT-4 \cr
  H86-70,SMC\_OGLE21          &  0:43:44 & -72:58:36  &  C  & 39 & 27 &  50 &  30 &mT-4  \cr
                                                    
  BS27,SMC\_OGLE177           &  0:44:55 & -73:10:27  &  C  & 24 & 21 &  80 & 19 &mP-14,in H86-72 \cr
  SMC-N10,L61-60,SMC-DEM11,  &  0:44:56 & -73:10:11  & NC  & 21 & 21 &  -  &  &mP-14,in H86-72 \& MA85 \cr

  NGC248n,SMC-N13B,L61-67n   &  0:45:24 & -73:22:34  & NA  & 42 & 36 & 110 & 21  &mP-15 \& SMC-DEM16n,ESO29EN8n,MA101,SMC\_OGLE26n \cr
  NGC248s,SMC-N13A,L61-67s   &  0:45:26 & -73:23:04  & NC  & 42 & 33 & 150 &  &mP-15 \& SMC-DEM16s,ESO29EN8s,MA103,SMC\_OGLE26s \cr
                      
  B39,SMC\_OGLE27          &  0:45:26 & -73:28:53 &   C & 33 & 33 &    -  & 15   &mP-16   \cr
  BS30                    &  0:45:30 & -73:29:06 &   C & 24 & 24 &    -  &  &mP-16   \cr 
                                                                                                         
  B36                     &  0:45:44 & -72:50:35 &   C & 42 & 36 &   140 & 10 &mP-17   \cr
  SMC\_OGLE31              &  0:45:51 & -72:50:25 &  CA & 33 & 27 &    90 &  &mP-17, not B36 \cr     
                                                  
  B38                     &  0:45:54 & -72:36:08 &  C  & 18 & 18 & -  & 38 &mP-18   \cr
  H86-79                  &  0:45:58 & -72:35:36 &  C  & 27 & 27 & -  &  &mP-18   \cr     
                                                   
  H86-76,SMC\_OGLE182      &  0:46:02 & -73:23:44 &    C  & 27 & 27 & -  &  &mT-5,in SMC-DEM21 \cr
  H86-78n,SMC\_OGLE33n     &  0:46:12 & -73:23:27 &    CN & 27 & 27 & -  & 16 &mT-5,in SMC-N16 \cr
  H86-78s,SMC\_OGLE33s     &  0:46:12 & -73:23:39 &    CN & 27 & 24 &  60 &  16 &mT-5,in SMC-N16  \cr     
                                    
  L31,SMC\_OGLE36          &  0:46:35 & -72:44:32     &  C  & 66 & 51 &  - &  &mT-6 \cr
  H86-83,SMC\_OGLE35       &  0:46:34 & -72:46:26     &  C  &42 & 42 &  - & 30 &mT-6 \cr
  H86-84,SMC\_OGLE185      &  0:46:34 & -72:45:56     &  C  &24 & 24 & - &  30 &mT-6  \cr
                                                    
  H86-86,SMC\_OGLE40       &  0:47:01 & -73:23:35     &  C  &48 & 39 &  110 &  71 &mP-19,in H-A9 \cr
  H86-87,SMC\_OGLE187      &  0:47:06 & -73:22:17     &  C  &48 & 42 &  90 &  &mP-19,in H-A9 \cr 
                        
  H86-95                  &  0:47:37 & -73:00:51    & CA &21 & 12 & 100 & 30 &mP-20   \cr
  H86-96                  &  0:47:44 & -73:00:46    &  C  &18 & 18 & -  &  &mP-20   \cr

  BS35,SMC\_OGLE42         &  0:47:50 & -73:28:42     &  C  &42 & 42 & -  & 55 &mP-21   \cr
  K25,L35,SMC\_OGLE45      &  0:48:01 & -73:29:10     &  C  &72 & 72 & -  &  &mP-21   \cr  
                                                    
  MA205                      &  0:48:07 & -73:14:49 &     NC &  15 & 15 & -  &  37 &mT-7 \cr
  SMC-N25,L61-106,SMC-DEM38, &  0:48:09 & -73:14:19 &    NA & 51 & 51 & -  &  &mT-7 \& MA208,SMC\_OGLE189 \cr
  SMC-N26,L61-107,MA206      &  0:48:08 & -73:14:53 &    NC & 27 & 27 & - &  37 &mT-7   \cr 
                                                  
  H86-99,SMC\_OGLE190      &  0:48:13 & -72:47:35    & CA &39 & 39 & -  & 27  &mP-22   \cr
  H86-100,SMC\_OGLE191     &  0:48:20 & -72:47:42    & CA &45 & 45 & -  &  &mP-22   \cr    
                                                  
  B50                     &  0:49:02 & -73:21:44     &  C  & 33 & 33 & -  & 66 &mT-8 \cr
  BS41,SMC\_OGLE194        &  0:49:06 & -73:21:10     &  C  & 33 & 33 & -  &  &mT-8 \cr
  L39,SMC\_OGLE54          &  0:49:18 & -73:22:20     &  C  &42 & 33 & 170 & 66 &mT-8,in BS43 \cr

  SMC-N33,L61-138,MA297   &  0:49:29 & -73:26:33    &  NC &  18 & 15 &  80 & 12 &mP-23,in SMC-DEM44 \cr
  MA301                   &  0:49:30 & -73:26:23    &  NC & 21 & 18 &  80 &  &mP-23,in SMC-DEM44 \cr 
                                        
  SMC\_OGLE56              &  0:49:36 & -72:50:13    & CA &48 & 36 & 100 & 70 &mT-9,in SMC-DEM46e \cr
  H86-110                 &  0:49:44 & -72:51:14    & CA &45 & 33 & 160 & 70 &mT-9 \cr
  H86-109,SMC\_OGLE58      &  0:49:45 & -72:51:58     &  C  &27 & 27 & -  &  &mT-9 \cr  

  MA317                      &  0:49:42 & -73:10:37  &   NC & 18 & 15 & 20 & 20 &mP-24   \cr 
  SMC-N34,L61-142,SMC-DEM50, &  0:49:46 & -73:10:25  &   NC & 39 & 27 & 120 &  &mP-24  \& MA322   \cr  
                                         
  B53,SMC\_OGLE197         &  0:50:04 & -73:23:04     &  C  &57 & 57 & -  &  80 &mP-25   \cr
  B55,SMC\_OGLE60          &  0:50:22 & -73:23:16     &  C  &42 & 36 & 110 &  &mP-25   \cr

  SMC\_OGLE199             &  0:50:15 & -73:03:15    & CA &15 & 15 & -  & 35 &mP-26, in? sup? SMC-DEM51 \cr
  BS45,SMC\_OGLE59         &  0:50:16 & -73:02:00    & CA & 60 & 54 & 70 &  &mP-26,in SMC-DEM51  \cr 
                                         
  H86-106w                &  0:50:31 & -73:20:11     &  C  &30 & 24 & 90 & 23 &mP-27,in SMC-DEM52 \cr
  H86-106e                &  0:50:37 & -73:20:11     &  C  &33 & 27 & 80 &  &mP-27,in SMC-DEM52 \cr

  H86-115,SMC\_OGLE63      &  0:50:37 & -73:03:28  &   AC & 96 & 72 &  40 & 77 &mP-28,in SMC-DEM51  \cr
  SMC\_OGLE65              &  0:50:55 & -73:03:27  &    C  &39 & 39 & -  &  &mP-28   \cr 

  BS46,SMC\_OGLE200        &  0:50:39 & -72:58:44     &  C  &30 & 27 &  60 & 47 &mP-29,in L40 \cr
  H86-116,SMC\_OGLE64      &  0:50:40 & -72:57:55     &  C  &30 & 30 &  -  &  &mP-29,in L40 \cr 

  BS48,SMC\_OGLE201        &  0:50:42 & -73:23:49    &  AC & 51 & 33 &  80 & 46 &mP-30,in SMC-DEM53 \cr 
  H86-108,MA401           &  0:50:53 & -73:24:22    &  NA & 60 & 60 &  -  &  &mP-30,in SMC-DEM53 \cr
  
   B59,L61-183,MA488,      &  0:51:44 & -72:50:25    & CN & 48 & 36 &  80 & 26 &mP-31 \& SMC\_OGLE73   \cr 
 
\hline
\end{tabular}
\end{scriptsize}
\label{tab:catalog}
\end{table*}

\setcounter{table}{0}
\begin{table*}
\caption[]{ Continued.}
\begin{scriptsize}
\renewcommand{\tabcolsep}{0.9mm}
\begin{tabular}{l r r r r r r r l}
\hline
\hline
Name & RA(2000) & Dec(2000)& T & Dmax & Dmin & PA & Separ. & Comments \cr
     & h : m : s &$^{o}$ : ' : ''&  & '' & '' & $^{o}$& '' & \cr 
(1)  & (2)   & (3)       &(4) & (5) & (6) & (7) & (8) & (9) \cr 
\hline
                                              
  SMC-N46,L61-184,SMC-DEM62, &  0:51:47 & -72:50:47 & NC & 39 & 39 &  -  &  &mP-31  \& MA498  \cr 
   
  H86-120                 &  0:51:46 & -73:28:01     &  C  & 21 & 21 & -  & 43 &mT-10,in SMC-DEM70s \cr
  BS53                    &  0:51:49 & -73:28:38     & A & 39 & 33 & 150 &  &mT-10,in SMC-DEM70s \cr
  H86-122                 &  0:51:58 & -73:27:41     &  C  & 27 & 27 & -  &  43 &mT-10,in SMC-DEM70s \cr     
                                     
  BS56,SMC\_OGLE77         &  0:52:13 & -73:00:12     &  C & 42 & 33 &  90 & 56 & mP-32  \cr
  H86-130,SMC\_OGLE78      &  0:52:17 & -73:01:04     &  C & 45 & 36 &   0 &  & mP-32  \cr  
                                                     
  B64,SMC\_OGLE210         &  0:52:30 & -73:02:59     &  C  &42 & 42 &  -  & 56 &mP-33,in H-A29 \cr
  BS57,SMC\_OGLE211        &  0:52:32 & -73:02:10     &  C  &39 & 27 &  60 &  &mP-33,in H-A29  \cr         
                                  
  H86-134w,SMC\_OGLE212    &  0:52:45 & -72:59:24     &  C  &30 & 30 & -  & 19 &mT-11,in H-A30 \cr
  B65,SMC\_OGLE83          &  0:52:44 & -72:58:48     &  C  &45 & 45 & -  &  &mT-11 \cr
  H86-134e,SMC\_OGLE213    &  0:52:48 & -72:59:22     &  C  &30 & 27 &  0 & 19 &mT-11,in H-A30 \cr
 
  BS61                    &  0:52:43 & -73:01:45    & CA &36 & 27 &  80 & 42 &mP-34,in H-A29 \cr
  BS255                   &  0:52:52 & -73:01:45    &  C  &24 & 18 & 70 &  &mP-34   \cr 
 
  BS63,SMC\_OGLE84            &  0:52:47 & -73:24:25     &  C  &30 & 24 & 150 & 19 &mT-12,in SMC-DEM73 \cr
  B67,SMC\_OGLE87             &  0:52:49 & -73:24:43     &  C  &39 & 30 & 110 &   &mT-12,in SMC-DEM73 \cr
  NGC294,L47,ESO29SC22,      &  0:53:06 & -73:22:49     &  C  &102 & 102 &  -  &  & mT-12 \& SMC\_OGLE90  \cr

  H86-140,SMC\_OGLE214     &  0:53:09 & -72:49:58     &  C  & 27 & 24 &  50 & 15&mP-35   \cr    
  H86-139                 &  0:53:11 & -72:50:05     &  C  & 21 & 18 &  40 &  &mP-35   \cr 

  BS67                    &  0:53:32 & -73:21:03    & AC &39 & 39 & -  & 51 &mP-36   \cr   
  BS68,SMC\_OGLE95         &  0:53:42 & -73:21:32    & CA &54 & 45 &130 &  &mP-36   \cr
                                                     
  B72                     &  0:53:26 & -72:40:57     &  C  &72 & 72 &  -  &  &m6,in H-A35 \cr
  H86-143,SMC\_OGLE93      &  0:53:31 & -72:40:04     &  C  & 48 & 48 &  -  &  &m6,in H-A35  \cr
  BS257                   &  0:53:36 & -72:38:30     & AC  &48 & 39 &  0 &  &m6 \cr    

  SMC-N52A,L61-243,       &  0:53:40 & -72:39:35     & NC & 30 & 30 & -  & 20&m6,in H-A35 \& SMC-DEM77sw,MA696,SMC\_OGLE94 \cr
  SMC-N52B,L61-244,B73,   &  0:53:42 & -72:39:15   &  NC & 30 & 30 & -  &  20 &m6,in H-A35 \& SMC-DEM77ne,MA699,SMC\_OGLE96 \cr
  H86-148                 &  0:53:55 & -72:40:08    &  C  &30 & 30 & -  &  &m6,in H-A35  \cr

  BS69,SMC\_OGLE217        &  0:53:56 & -72:51:24    & CA &36 & 24 &  45 &  75 &mP-37   \cr
  BS72,SMC\_OGLE97         &  0:54:11 & -72:51:54    & CA &45 & 36 &  20 &  &mP-37   \cr 
                                                       
  B78		    &  0:54:45 & -72:07:46     &  C  & 66 & 48 & 110 & 56&mP-38,in H-A37 \cr
  L51,ESO51SC7            &  0:54:54 & -72:06:46     &  C  & 60 & 45 & 170 &  &mP-38,in H-A37 \cr    
                                        
  H86-159,SMC\_OGLE102     &  0:55:12 & -72:41:00     &  C  &30 & 24 & 130 & 45 &mP-39,in BS260 \cr
  H86-160                 &  0:55:21 & -72:40:10     &  C  &24 & 24 &  -  &  &mP-39,in BS260 \cr   
                                            
  BS81,SMC\_OGLE223        &  0:56:26 & -72:29:45     &  C  &36 & 33 &  0 & 43 &mP-40,in H-A40 \cr
  H86-172,SMC\_OGLE108     &  0:56:34 & -72:30:08     &  C  &33 & 33 & -  &  &mP-40,in H-A40 \cr  
                                                                                                  
  H86-175,SMC\_OGLE227     &  0:57:50 & -72:26:24     &  C  &24 & 24 & -  & 38 &mP-41   \cr 
  H86-179,SMC\_OGLE112     &  0:57:57 & -72:26:42     &  C  &24 & 24 & -  &  &mP-41   \cr 

  H86-177,SMC\_OGLE226     &  0:57:50 & -72:30:29     &  C  &45 & 45 & -  & 46 &mP-42,in B-OB13 \cr
  H86-176                 &  0:57:53 & -72:29:48     &  C  &36 & 30 & 90 &  &mP-42   \cr 
                                                     
  SMC-N62,SMC-DEM93          &  0:57:56 & -72:39:26    &  NA & 72 &  72 & -  & 75 &mT-13,in H-A42  \cr  
  SMC-N63,L61-331,SMC-DEM94, &  0:58:16 & -72:38:47    & NA & 36 & 36 & -  &  &mT-13,in H-A42 \& MA1065,SMC\_OGLE113 \cr
  SMC-N64A,L61-335,SMC-DEM95        &  0:58:26 & -72:39:57    &  NC & 48 & 39 &  70 & 75 &mT-13,in SMC-N64 \& H86-182,MA1071,SMC\_OGLE114  \cr

  BS269                   &  0:58:19 & -72:13:10    & CA & 24 & 18 &  60 & 35 &mP-43   \cr
  BS270                   &  0:58:23 & -72:12:43    & CA & 33 & 30 & 130 &  &mP-43  \cr   
                                                     
  BS271                   &  0:58:37 & -72:13:27    & NC & 39 & 30 & 30 & 35 &mP-44,in SMC-DEM98 \cr
  BS272,SMC\_OGLE229       &  0:58:38 & -72:14:04    & NC & 39 & 39 & -  &  &mP-44,in SMC-DEM98 \cr   
                                        
  BS93                    &  0:59:36 & -71:44:13     &  C  &24 & 21 &  10 & 17 &mT-14,in SMC-DEM105 \cr
  B97                     &  0:59:37 & -71:44:40     &  C  &30 & 30 &  -  &  &mT-14,in SMC-DEM105 \cr
  BS273                   &  0:59:40 & -71:44:34     & AC  &30 & 27 & 120 &  17 &mT-14,in SMC-DEM105  \cr    
                                   
  IC1611,K40,L61,ESO29SC27,  &  0:59:48 & -72:20:02     &  C  & 90 & 90 & -  &  &m4-1 \& SMC\_OGLE118 \cr
  H86-186,SMC\_OGLE119        &  0:59:57 & -72:22:24     &  C  &36 & 36 & -  &  29 &m4-1,att SMC-DEM114 \cr
  IC1612,K41,L62,ESO29SC28,  &  1:00:01 & -72:22:08     &  C  &72 & 48 & 20 & 29 &m4-1,att SMC-DEM114 \& SMC\_OGLE120  \cr 
  K42,L63,SMC\_OGLE124     &  1:00:34 & -72:21:56     &  C  & 51 & 51 &   -  &  &m4-1,att SMC-DEM114 \cr
 
  B98sw                   &  1:00:21 & -73:52:51     &  C  & 36 & 27 & 40 & 33 &mP-45   \cr
  B98ne                   &  1:00:28 & -73:52:32     &  C  & 33 & 27 & 130 &  &mP-45   \cr 

  H86-189,SMC\_OGLE123     &  1:00:33 & -72:14:23     &  C  &24 & 24 & -  & 68 &mP-46   \cr
  H86-190,SMC\_OGLE230     &  1:00:33 & -72:15:31     &  C  &24 & 24 & -  &  &mP-46   \cr             
                                                     
  H86-191,SMC\_OGLE231     &  1:00:58 & -72:32:25     &  C  &48 & 48 & -  & 81  &mP-47,in? SMC-DEM114 \cr
  H86-194,SMC\_OGLE232     &  1:01:14 & -72:33:03     &  C  &51 & 51 & -  &  &mP-47,in? SMC-DEM114  \cr 

  BS102                   &  1:01:14 & -73:47:45     &  C  &30 & 21 & 100 & 42 &mP-48   \cr
  HW44                    &  1:01:22 & -73:47:15     &  C  &45 & 45 & -  &  &mP-48   \cr
                                                      
  B110                    &  1:02:11 & -72:00:11     &  C  &33 & 27 & 70 & 65 &mP-49, br*in,in H-A49 \cr
  B112                    &  1:02:23 & -72:00:11     &  C  &51 & 45 & 60 &  &mP-49, br*in,in H-A49 \cr 
                                        
  K45w,L69w               &  1:02:45 & -73:44:19     &  C  &33 & 27 & 40 & 47 &mP-50   \cr
  K45e,L69w               &  1:02:49 & -73:44:25     &  C  &24 & 24 & -  &  &mP-50   \cr  

  NGC376,K49,L72,ESO29SC29,&  1:03:53 & -72:49:34     &  C  & 108 & 108 &  -  & 75 &mP-51   \& SMC\_OGLE139 \cr
  BS114,SMC\_OGLE235        &  1:03:59 & -72:48:18     & AC  & 42 & 30 & 110 &  &mP-51   \cr 
                                                      
  SMC\_OGLE138             &  1:03:53 & -72:06:11    & CA &36 & 36 & -  & 82 &mP-52,in H-A53 \cr
  SMC\_OGLE144,SMC\_OGLE236 &  1:04:05 & -72:07:15    & CA &36 & 36 & -  &  &mP-52   \cr 
                                                      
  BS122                   &  1:04:18 & -73:10:21     &  C  &21 & 21 & -  & 26  &mP-53   \cr
  B119                    &  1:04:19 & -73:09:54     &  C  &36 & 36 & -  &  &mP-53   \cr    
                                                  
  SMC-N78A,L61-438,MA1512 &  1:05:04 & -71:59:01     & NC & 24 & 21 & 140 & 23 &m5,in NGC395 \cr
  SMC-N78B,L61-439,       &  1:05:04 & -71:59:25     & NC & 24 & 18 & 100 &  &m5,in NGC395 \& MA1508/1514,SMC\_OGLE145 \cr 
  MA1520,SMC\_OGLE147      &  1:05:08 & -71:59:45    & NC & 30 & 27 & 130 &  23 &m5,in NGC395 \cr
  SMC-N78D,SMC-DEM127     &  1:05:11 & -71:58:28    & NA & 54 & 54 &  -  &  &m5,att SMC-N78 \cr
  SMC\_OGLE146             &  1:05:13 & -71 59 42    & NA & 27 & 27 &  -  &  &m5,in NGC395 \cr   

  BS130                   &  1:05:56 & -72:04:11    & A  & 57 & 36 & 140 &  30&mP-54,att SMC-N78  \cr
  BS132                   &  1:06:01 & -72:03:37    & CA & 42 & 33 & 150 &  &mP-54,att SMC-N78  \cr             
                              
  BS133                   &  1:06:23 & -71:55:12    &  A & 39 & 30  & 30 &  &mT-15,in BS134 \cr
  BS135                   &  1:06:40 & -71:55:13    & CA & 39 & 39  & -  & 47 &mT-15,in BS134 \cr
  BS136                   &  1:06:48 & -71:54:55    & CA & 42 & 42  & -  & 47 &mT-15,in SMC-DEM132 \cr          
                                                   
  B134                    &  1:09:01 & -73:12:24   & CA & 48 & 33 & 80 &  &m4-2,in H-A60 \cr
  BS142                   &  1:09:07 & -73:12:01   &  C & 24 & 21 & 70 & 31 &m4-2,in H-A60 \cr
  IC1644,SMC-N81,L61-481, &  1:09:12 & -73:11:42   & NC & 48 & 39 & 40 & 31 &m4-2,in H-A60 \& ESO29EN35,MA1688/1687 \cr 
  B135                    &  1:09:19 & -73:11:15   &  C & 33 & 24 & 60 &  &m4-2,in H-A60 \cr
 
  NGC422,K62,L87,ESO51SC22 &  1:09:25 & -71:46:00     &  C  & 60 & 60 & -  & 64 &mP-55   \cr 
  IC1641,HW62,ESO51SC21    &  1:09:39 & -71:46:07     &  C  & 45 & 39 & 40 &  &mP-55   \cr                     
 
  HW71nw                  &  1:15:30 & -72:22:36     &  C  & 18 & 18 &  -  & 17 &mP-56   \cr
  HW71se                  &  1:15:33 & -72:22:50     &  C  & 39 & 33 & 170 &  &mP-56   \cr 
\hline
\end{tabular}
\end{scriptsize}
\end{table*}

\section{The Isophotal Atlas}

For isophotal analysis purposes we extracted digitized images of pairs and
multiplets from the DSS. The plates are from the
SERC Southern Sky Survey and include IIIa-J long (3600s), V band medium (1200s)
and V band short (300s) exposures. The PDS pixel values correspond to
photographic density measures from the original plates, and are not
calibrated. The images were processed with the IRAF package at the Instituto de
F\'{\i}sica - UFRGS, applying a 2-d Gaussian filter to smooth out individual
stars, producing isodensity maps. 

 We show the isophotal atlas of SMC star cluster pairs and multiplets in
Figs.~\ref{fig:atlas1} throughout \ref{fig:atlas7}, where   morphological
evidence of interactions can be searched for. In each panel we point  members and
 provide designations.

\subsection{Isophotal Distortions}

Cluster pairs and multiplets with evidence of physical interaction are marked in
column 6 of Table~\ref{tab:interact}. Their isophotal maps show features such as
isophotal distortions, common envelope, isophotal twistings, etc. We classified
these isophotal maps according to the following criteria: (i) isophotes of the
members are detached ({\bf d}), but showing isophotal distortions; (ii)
isophotes of the members are connected by a ``bridge'' ({\bf b}); (iii) the
members are embedded in the same isophotal envelope ({\bf e}). These
classification criteria are an important tool for a selection of interacting
cluster pair candidates, since such isophotal morphologies are predominant in
N-body model encounters (Rodrigues et al. 1994, ODB98). The
classifications above were made for objects not disturbed by gas emission.
 It is worth noting that both bridges and envelopes imply a common outer isophotal, however 
the former includes a dimension which is considerably smaller than the smaller cluster diameter. 
The pair B59/SMC-N46 would be an  envelope limiting case (Fig.~\ref{fig:atlas4}). 

 As examples, the pair  H86-186/IC1612 (Fig.~\ref{fig:atlas5}) is embedded in the
 same isophotal envelope and it is classified as {\bf e}. The pair
 H86-140/H86-139 (Fig.~\ref{fig:atlas4}) shows a bridge linking its members and
 is classificated as {\bf b}. In Fig.~\ref{fig:atlas6} the pair NGC376/BS114
 shows isophotal distortions, however the outer isophotes of the components are
 not connected and it is thus classified as {\bf d}.

We detected relevant isophotal features for about 25\% of the sample distributed
as follows: 6 candidate cluster systems with common envelope, 7 with bridge and 5 detached cases.

High isophotal densities such as those observed for B78/L51 (Fig.~\ref{fig:atlas5}) are related to the
high surface brightness which often occurs in blue clusters.

 It is important to note that non relevant isophotal features can appear in the maps caused by bright stars
or background fluctuations. By means of previous inspections in the original images we verified the
non relevant objects. For example, the compact object in the lower right corner of the B39/BS30 isophotal
map (Fig.~\ref{fig:atlas2}) is a bright star, however the compact one to the right of the main cluster
can be a star or a compact cluster, as seen in the original form. Higher resolution images would be
necessary to check the presence of a third cluster.

\begin{table*}
\caption[]{Age, Reddening and Morphological Classification}
\begin{scriptsize}
\renewcommand{\tabcolsep}{0.9mm}
\begin{tabular}{l r r r r r l}
\hline
\hline
Name & Age & E(B-V)& Age$_{OGLE}$ & E(B-V)$_{OGLE}$ & Morphologies & Commments  \cr
     & Myr &       &  Myr     &             & 		  &  	  \cr
(1)  & (2) &  (3)  &   (4)    &      (5)    &    (6)  	  &  (7)  \cr
\hline

  BS3 			  &         &   	& 	  & 	 &b   &   mT-1 outside OGLEII	\cr 
  H86-2 		  & 	 & 	 &	 & 	 & 	b &    mT-1 outside OGLEII	\cr 
  BS4 			  & 	 & 	 & 	&	&        &	mT-1 outside OGLEII  \cr

  H86-38                  &251$\pm$25 &0.10&&&&mP-5   \cr
  BS10                    &$>$560 &0.10&&&&mP-5    \cr

  B15                     &$>$1000 &0.07&&&&mP-8    \cr
  H86-57                  &$>$1000 &0.07&&&&mP-8    \cr

  NGC220,SMC\_OGLE8    &65$\pm$13   &0.07&100$\pm$23&0.07&d& mT-2   \cr
  NGC222,SMC\_OGLE9    &70$\pm$7  &0.07&100$\pm$23&0.07&d& mT-2   \cr
  B23,SMC\_OGLE170         & $<$100 &0.07&&&d& mT-2  \cr

  NGC231,SMC\_OGLE11   &65$\pm$8  &0.12&79$\pm$18&0.08&&mP-10    \cr 
  BS15                    &70$\pm$7  &0.10&&&&mP-10   \cr

  B29                     & 150$\pm$70  & 0.10 &&&&mP-12   \cr 
  HW16,SMC\_OGLE13         &$<$20  &0.07&20$\pm$15&0.05&&mP-12\cr
                                        
  NGC241,SMC\_OGLE17     &55$\pm$5  &0.12&79$\pm$18&0.10&b&mP-13  \cr
  NGC242,SMC\_OGLE18    &65$\pm$10  &0.10&79$\pm$18&0.10&b&mP-13  \cr
                                   
  B31,SMC\_OGLE19,SMC\_OGLE175 &280$\pm$30  &0.10&400$\pm$92&0.08&&mT-4 \cr
  BS20,SMC\_OGLE20            &450$\pm$50  &0.08&400$\pm$92&0.08&&mT-4 \cr
  H86-70,SMC\_OGLE21          &450$\pm$50  &0.07&&&&mT-4  \cr
                                                    
  BS27,SMC\_OGLE177           &$<$25  &0.18&79$\pm$37&0.08&&mP-14   \cr
  SMC-N10		      &HII Region  &&&&&mP-14 \cr

  NGC248n,SMC\_OGLE26n    &HII Region  &&&&&mP-15  \cr
  NGC248s,SMC\_OGLE26s    &HII Region  &&&&&mP-15  \cr
                      
  B39,SMC\_OGLE27          &450$\pm$50 &0.07&&&b&mP-16  \cr
  BS30                    &450$\pm$50  &0.07&&&b&mP-16   \cr 
                                                                                                         
  B36                     &315$\pm$50  &0.03&&&&mP-17  \cr
  SMC\_OGLE31              &450$\pm$50  &0.03&&&&mP-17  \cr                                                 
                                                   
  H86-76,SMC\_OGLE182      & 200$\pm$20 &0.14&&&&mT-5  \cr
  H86-78n,SMC\_OGLE33n     & $<$25 &0.14&16$\pm$9&0.15&&mT-5  \cr
  H86-78s,SMC\_OGLE33s     &$<$30  &0.14&16$\pm$9&0.15&&mT-5  \cr     
                                    
  L31,SMC\_OGLE36          &250$\pm$50  &0.10&&&&mT-6  \cr
  H86-83,SMC\_OGLE35       &180$\pm$50  &0.10&&&&mT-6  \cr
  H86-84,SMC\_OGLE185      &250$\pm$50  &0.10&&&&mT-6  \cr
                                                    
  H86-86,SMC\_OGLE40       &350$\pm$50  &0.03&&&&mP-19 \cr
  H86-87,SMC\_OGLE187      &140$\pm$20  &0.07&158$\pm$36&0.04&&mP-19 \cr 
                        
  BS35,SMC\_OGLE42         &400$\pm$100 &0.03&&& &mP-21   \cr
  K25,SMC\_OGLE45      &250$\pm$50  &0.07&250$\pm$58&0.07&&mP-21   \cr  
                                                    
  MA205                      &HII Region  &&&&b&mT-7 \cr
  SMC-N25,SMC\_OGLE189 	&HII Region  &&&&b&mT-7  \cr
  SMC-N26		      &HII Region  &&&&&mT-7   \cr 
                                                  
  H86-99,SMC\_OGLE190      &225$\pm$50  &0.14&&&d&mP-22   \cr
  H86-100,SMC\_OGLE191     &200$\pm$50  & 0.14 &&&d&mP-22   \cr    
                                                  
  B50                     &$<$30  &0.10&&&&mT-8 \cr
  BS41,SMC\_OGLE194        &70$\pm$30  &0.10&79$\pm$18 & 0.07 &&mT-8 \cr
  L39,SMC\_OGLE54          &80$\pm$20  &0.10 & 100$\pm$23 & 0.10 &&mT-8 \cr

  SMC-N33 		  & HII Region &&&&e&mP-23 \cr
  MA301                   &HII Region  &&&&e&mP-23 \cr 
                                        
  SMC\_OGLE56              &115$\pm$15  &0.18&&&&mT-9 \cr
  H86-110                 &$<$20  &0.18&&&&mT-9 \cr
  H86-109,SMC\_OGLE58      &180$\pm$20  &0.18&200$\pm$45&0.17&&mT-9 \cr  

  MA317                      &HII Region  &&&&b&mP-24   \cr 
  SMC-N34			 &HII Region  &&&&b&mP-24     \cr  
                                         
  B53,SMC\_OGLE197         &200$\pm$50  &0.07&250$\pm$55&0.08&&mP-25   \cr
  B55,SMC\_OGLE60          &160$\pm$30  &0.07&250$\pm$55&0.08&&mP-25   \cr

  SMC\_OGLE199             & --  & -- &&&&mP-26  \cr
  BS45,SMC\_OGLE59         & 65$\pm$30  & 0.10 & 63$\pm$14 & 0.10 &&mP-26  \cr 
                                         
  H86-106w                &200$\pm$50  &0.10&&&e&mP-27 \cr
  H86-106e                & $<$30/200$\pm$50$^{1}$  &0.12/0.10&&&e&mP-27  \cr

  H86-115,SMC\_OGLE63      & --  & -- &&&&mP-28   \cr
  SMC\_OGLE65              &200$\pm$20  &0.10&&&&mP-28   \cr 

  BS46,SMC\_OGLE200        &80$\pm$10  &0.07&100$\pm$23&0.06&b&mP-29 \cr
  H86-116,SMC\_OGLE64      &125$\pm$15  &0.10&126$\pm$29&0.10&b&mP-29 \cr 

  BS48,SMC\_OGLE201        &160$\pm$20  &0.10&&&&mP-30 \cr 
  H86-108,MA401           &$<$30  &0.12&&&&mP-30 \cr
  
   B59,SMC\_OGLE73      &100$\pm$15  &0.14&158$\pm$75&0.11&e&mP-31  \cr 
   SMC-N46		 &HII Region  &&&&e&mP-31   \cr 

\hline
\end{tabular}
\end{scriptsize}
\label{tab:interact}
\end{table*}

\setcounter{table}{1}
\begin{table*}
\caption[]{Continued.}
\begin{scriptsize}
\renewcommand{\tabcolsep}{0.9mm}
\begin{tabular}{l r r r r r l}
\hline
\hline
Name & Age & E(B-V)& Age$_{OGLE}$ & E(B-V)$_{OGLE}$ & Morphologies & Comments  \cr
     & Myr &       &  Myr     &             & 		  &  	  \cr
(1)  & (2) &  (3)  &   (4)    &      (5)    &    (6)  	  &  (7)  \cr
\hline

  BS56,SMC\_OGLE77         &140$\pm$20  &0.07&79$\pm$38&0.08&d& mP-32  \cr
  H86-130,SMC\_OGLE78      &65$\pm$8  &0.12&79$\pm$18&0.08&d& mP-32  \cr  
                                                     
  B64,SMC\_OGLE210         &$<$10  &0.12&158$\pm$75&0.07&&mP-33  \cr
  BS57,SMC\_OGLE211        &$<$10  &0.12&&&&mP-33  \cr         
                                  
  H86-134w,SMC\_OGLE212    & --  & -- &&&&mT-11 \cr
  B65,SMC\_OGLE83          &65$\pm$10  &0.10&63$\pm$30&0.09&&mT-11 \cr
  H86-134e,SMC\_OGLE213    & --  & -- &&&&mT-11 \cr
 
  BS61                    &250$\pm$50  &0.03&&&&mP-34 \cr
  BS255                   & --  & -- &&&&mP-34    \cr 
 
  BS63,SMC\_OGLE84            &450$\pm$50  &0.10&&&d&mT-12 \cr
  B67,SMC\_OGLE87             &450$\pm$50  &0.10&500$\pm$115&0.10&d&mT-12 \cr
  NGC294,SMC\_OGLE90    &300$\pm$50  &0.10&316$\pm$73&0.11&d& mT-12  \cr

  H86-140,SMC\_OGLE214     &55$\pm$25  &0.10&&&b&mP-35   \cr    
  H86-139                 & $<$30  &0.10&&&b&mP-35   \cr 

  BS67                    &$>$300  &0.07&&&&mP-36   \cr   
  BS68,SMC\_OGLE95         &500$\pm$100  &0.07&&&&mP-36   \cr
                                                     
  B72                     & 80$\pm$10  & 0.10 &&&&m6  \cr
  H86-143,SMC\_OGLE93      &200$\pm$25  &0.10&&&&m6   \cr
  BS257                   &50$\pm$30$^{1}$  & 0.10&&&&m6 \cr    
  SMC-N52A,SMC\_OGLE94      &HII Region  &&&&&m6 \cr
  SMC-N52B,SMC\_OGLE96    &HII Region  	 &&&&&m6 \cr
  H86-148                 &400$\pm$100  &0.10&&&&m6  \cr
                                                       
  H86-159,SMC\_OGLE102     &500$\pm$50  &0.10&&&&mP-39  \cr
  H86-160                 & --  & -- &&&&mP-39  \cr   
                                            
  BS81,SMC\_OGLE223        &80$\pm$20/250$\pm$50$^{1}$  &0.10&&&&mP-40 \cr
  H86-172,SMC\_OGLE108     &280$\pm$30  &0.07&&&&mP-40  \cr  
                                                                       
  H86-175,SMC\_OGLE227     &30$\pm$10  &0.10&&&&mP-41   \cr 
  H86-179,SMC\_OGLE112     &$<$30  &0.10 & 32$\pm$23 & 0.11 &&mP-41   \cr 

  H86-177,SMC\_OGLE226     &$<$30  &0.10&&&d&mP-42 \cr
  H86-176                 &$<$30  &0.10&&&d&mP-42  \cr 
                                                     
  SMC-N62	          &HII Region  &&&&&mT-13  \cr  
  SMC-N63,SMC\_OGLE113    &HII Region  &&&&&mT-13 \cr
  SMC-N64A,SMC\_OGLE114    &HII Region   &&&&&mT-13 \cr

  BS269                   & --  & -- &&&&mP-43   \cr
  BS270                   &$<$30  &0.10&&&&mP-43   \cr   
                                                     
  BS271                   & $<$30 &0.12&&&&mP-44 \cr
  BS272,SMC\_OGLE229       &$<$30  &0.08&79$\pm$18&0.05&&mP-44 \cr

  IC1611,SMC\_OGLE118	  & 100$\pm$20 & 0.07 & 160$\pm$37 & 0.08 &&m4-1   \cr
  H86-186,SMC\_OGLE119    & 180$\pm$20 & 0.07 &&&e&m4-1 \cr
  IC1612,SMC\_OGLE120	 & 100$\pm$50  & 0.10 & 50$\pm$24 & 0.07 &e&m4-1  \cr 
  K42,SMC\_OGLE124   	  &20$\pm$10  &0.10& 40$\pm$9 & 0.06 &&m4-1 \cr

  H86-189,SMC\_OGLE123     & 570$\pm$70  & 0.07 &&&&mP-46   \cr
  H86-190,SMC\_OGLE230     & 125$\pm$13 & 0.07 &32$\pm$23&0.08&&mP-46   \cr             
                                                     
  H86-191,SMC\_OGLE231     &200$\pm$25 &0.07&&&&mP-47 \cr
  H86-194,SMC\_OGLE232     &200$\pm$25  &0.07&&&&mP-47 \cr

  NGC376,SMC\_OGLE139	  &20$\pm$2  &0.10&32$\pm$7&0.07&e&mP-51    \cr
  BS114,SMC\_OGLE235        &250$\pm$25  &0.04&&&e&mP-51   \cr 
                                                      
  SMC\_OGLE138             & $<$30  & 0.04 & 25$\pm$20 & 0.04 &&mP-52 \cr
  SMC\_OGLE144,SMC\_OGLE236 & $<$30  & 0.07 & 40$\pm$19 & 0.05 &&mP-52  \cr

  SMC-N78A		 & HII Region &&&&&m5 \cr
  SMC-N78B,SMC\_OGLE145      & HII Region && 79$\pm$37 & 0.07 &&m5 \cr 
  MA1520,SMC\_OGLE147      &HII Region  && 12$\pm$9 & 0.06 &&m5\cr
  SMC-N78D	    	 &HII Region  &&&&&m5 \cr
  SMC\_OGLE146             & HII Region  && 20$\pm$15 & 0.06 &&m5 \cr   

  BS130                   &80$\pm$10 &0.10&&&&mP-54  \cr
  BS132                   & $<$30/200$\pm$25$^{1,2}$  & 0.10 &&&&mP-54 \cr             
                              
  HW71nw                  &  &&&&e&mP-56 outside OGLEII  \cr
  HW71se                  &  &&&&e&mP-56 outside OGLEII  \cr 
\hline
\end{tabular}
\end{scriptsize}
\begin{list}{}
\item Notes to Table~\ref{tab:interact}: 1- Age depends on membership of bright stars; 
2- Upper main sequence is possibly  underpopulated.  
\end{list}
\end{table*}

\begin{figure*}
%\resizebox{\hsize}{!}{\includegraphics{catal_1.gif}}
\caption{Isophotal atlas of SMC cluster pairs and multiplets.}
\label{fig:atlas1}
\end{figure*}

\begin{figure*}
%\resizebox{\hsize}{!}{\includegraphics{catal_2.gif}}
\caption{Isophotal atlas of SMC cluster pairs and multiplets.}
\label{fig:atlas2}
\end{figure*}

\begin{figure*}
%\resizebox{\hsize}{!}{\includegraphics{catal_3.gif}}
\caption{Isophotal atlas of SMC cluster pairs and multiplets.}
\label{fig:atlas3}
\end{figure*}

\begin{figure*}
%\resizebox{\hsize}{!}{\includegraphics{catal_4.gif}}
\caption{Isophotal atlas of SMC cluster pairs and multiplets.}
\label{fig:atlas4}
\end{figure*}

\begin{figure*}
%\resizebox{\hsize}{!}{\includegraphics{catal_5.gif}}
\caption{Isophotal atlas of SMC cluster pairs and multiplets.}
\label{fig:atlas5}
\end{figure*}

\begin{figure*}
%\resizebox{\hsize}{!}{\includegraphics{catal_6.gif}}
\caption{Isophotal atlas of SMC cluster pairs and multiplets.}
\label{fig:atlas6}
\end{figure*}

\begin{figure*}
%\resizebox{\hsize}{!}{\includegraphics{catal_7.gif}}
\caption{Isophotal atlas of SMC cluster pairs and multiplets.}
\label{fig:atlas7}
\end{figure*}

\section{Ages of cluster members}

Udalski et al. (1998) provided a BVI photometric survey of the SMC central $\approx$ 2.4
square degrees  with results for about
2.2 million stars. The data were collected during the OGLE-II microlensing
search project (Udalski et al. 1997). 

Pietrzy\'nski \& Udalski (1999b) studied colour-magnitude diagrams (CMDs)
extracted from the BVI database for 93 SMC star clusters, taking into account
neighbouring field extractions for comparisons. They estimated reddening values for
each cluster using: (i) the mean I band magnitude of red clump stars in the
cluster neighbourhood; (ii) the assumed extinction-free magnitude of the SMC's
red clump stars, I = 18.34 mag (Udalski 1998). They derived ages using Bertelli
et al.'s (1994) isochrones, adopting as a rule for the SMC a
metallicity Z = 0.004, and an absolute distance modulus (m-M) = 18.65.

Considering the OGLE-II angular coverage we conclude that out of the present 176
objects 133 are therein included. We obtained from the database V and I band
CMDs for these objects, using their coordinates and diameters to define a box
extraction.  Guided by DSS images of each system we selected representative
field regions to extract stars and construct CMDs for comparison purposes. Ages
and reddenings for the clusters were estimated by fitting the Padova
isochrones. We adopted Bertelli et al.'s (1994) isochrones rather than the
new isochrones by Girardi  et al.  (2000), since the previous set includes
younger ages. The isochrone age grid suitable for 4, 5, 6.3, 7.9, 10, 12, 16,
20, 25, 32, 40, 50, 63, 79, 100, 126, 160, 199, 251, 316, 400, 500, 630, 790 and
1000 Myr. The isochrone age range and resolution allow one to estimate errors
taking into account the stellar statistics in the cluster and field CMDs.  We
also assumed a SMC metallicity of Z = 0.004, a foreground galactic reddening of
E(B-V)$_{f}$ = 0.03 in the SMC direction and E(V-I)/E(B-V) =
1.31. Table~\ref{tab:interact} shows the ages, in column 2 and reddening values
(foreground plus SMC internal) in column 3 determined in the present work via
isochrone fitting. Ages and reddening values derived by Pietrzy\'nski \& Udalski
(1999b) are in column 4 and 5 respectively, when available for comparisons. In
our sample we could estimate CMD ages for 91 objects, 40 of them in common with
Pietrzy\'nski \& Udalski (1999b).  We conclude that despite the reddening method
and distance modulus differences between the two studies, there is good overall
agreement for the age determinations.  We also included ages for 21 objects
embedded in HII regions, classified as NA and NC (see column 4 of
Table~\ref{tab:catalog}, and also BS95). We assumed for them an age of 3 Myr.

\section{Discussion}

\begin{figure}
\resizebox{\hsize}{!}{\includegraphics{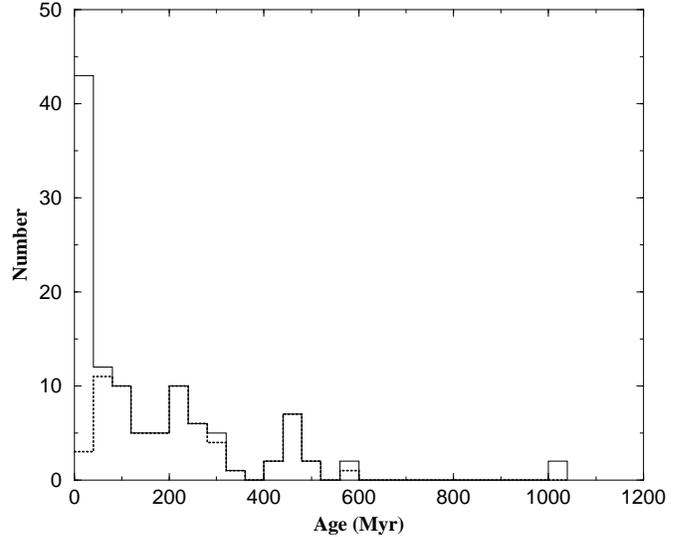}}
\caption{Age histograms: dotted lines represent clusters with age via CMDs; solid lines are for the same
sample as above plus embedded clusters in HII regions, assuming an age 3 Myr. In the latter histogram
lower or upper limits in Table~\ref{tab:interact} were also included by adopting them as  cluster ages.}
\label{fig:age}
\end{figure}

\begin{figure}
\resizebox{\hsize}{!}{\includegraphics{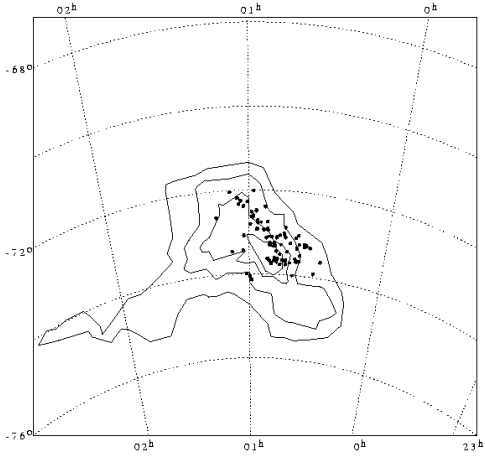}}
\caption{Angular distribution of the 176 objects in SMC cluster pairs and
multiplets together with HI contours 100, 200, 400 and 600 in units $10^{19}$
atoms $cm^{-2}$ from Mathewson \& Ford (1984).}
\label{fig:distr_SMC}
\end{figure}

\begin{figure}
\resizebox{\hsize}{!}{\includegraphics{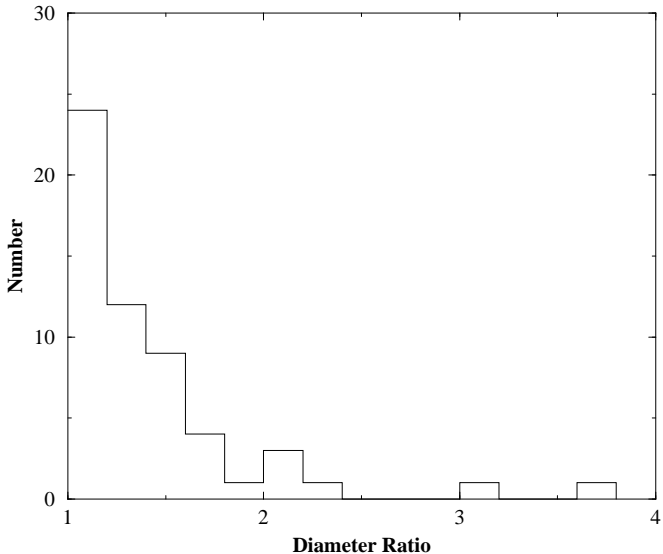}}
\caption{The observed diameter ratio distribution for SMC cluster pairs}
\label{fig:diam_SMC}
\end{figure}

\begin{figure}
\resizebox{\hsize}{!}{\includegraphics{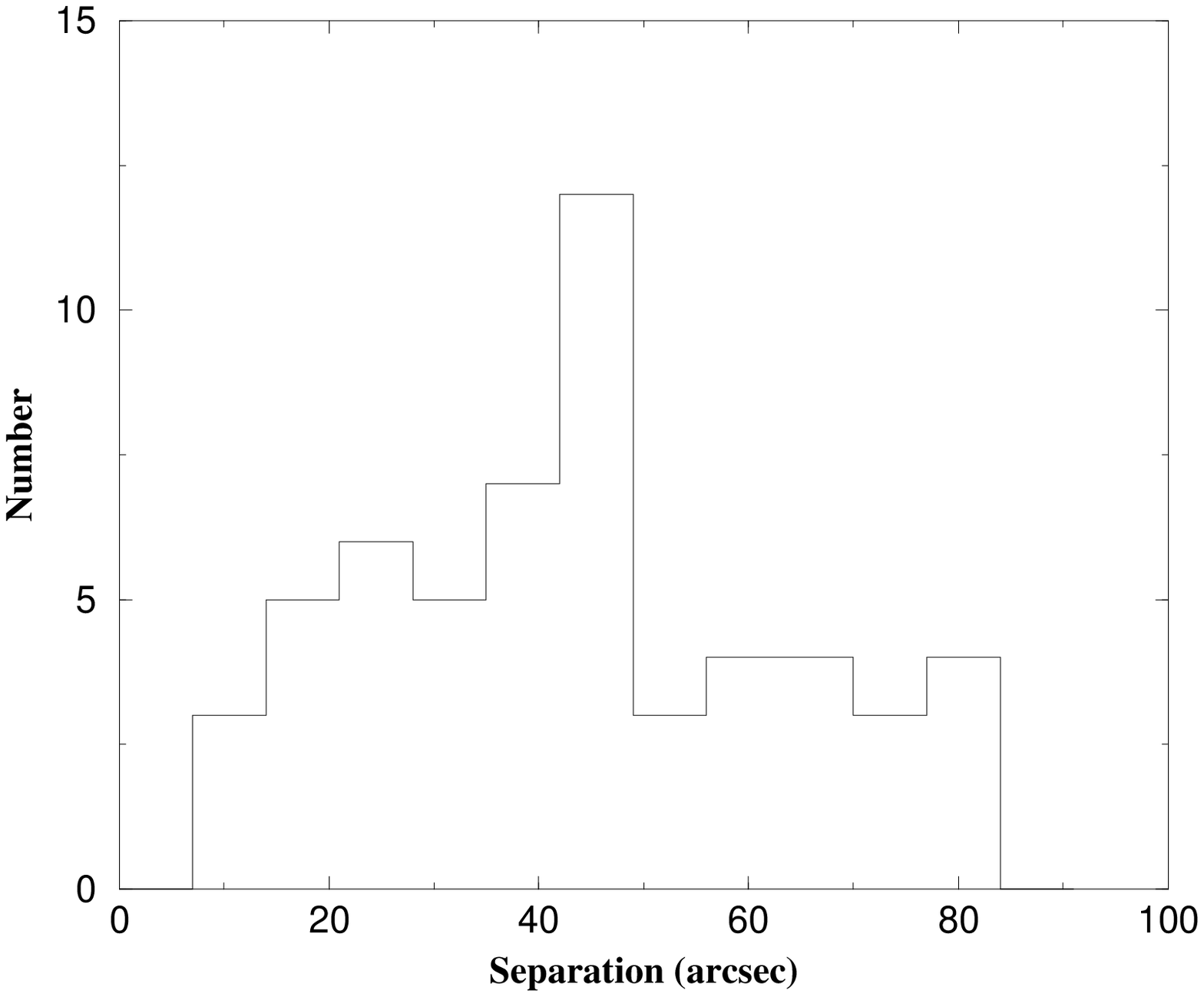}}
\caption{Centre-to-centre angular separation distribution for SMC cluster pairs}
\label{fig:separ_SMC}
\end{figure}

 We overlap two age histograms in Fig.~\ref{fig:age}. The dotted line histogram
  shows objects from Table~\ref{tab:interact} with ages via CMDs.  There are
  three peaks, at about 80, 220 and 450 Myr. The first two peaks appear to be
  present in Grebel et al.'s (2000) sample for 200 SMC clusters based on
  OGLE-II data, reported at t $\approx$ 100 and 200 Myr respectively. They
  suggested them as enhanced star formation epochs. Considering Pietrzy\'nski \&
  Udalski's (1999b) sample with CMD ages for 93 SMC clusters, no peak is seen at
  t = 80-100 Myr, having their histogram a maximum at the youngest bin. In their
  histogram there occurs a local maximum at t $\approx$ 250-300 Myr, but the
  statistics is low.  The third peak in the present study has no counterpart in
  both previous analyses, and it is probably an artifact from the low isochrone
  age step resolution in that range.

  The solid line histogram in Fig.~\ref{fig:age} shows additionally the embedded
  objects in HII regions for which we assumed an age of 3 Myr, and the objects
  from Table~\ref{tab:interact} with lower or upper limits by assuming them as
  the ages themselves. We can observe again a three peak distribution but the
  first peak is now shifted to the youngest bin similarly to Pietrzy\'nski \&
  Udalski's (1999b) histogram. This suggests that only the 200 Myr peak is
  relevant, being related to the SMC/LMC last encounter (Gardiner  et al.
  1994, Grebel et al.'s 2000). The maximum seen in the youngest bin is
  possibly related to the cluster formation/destruction rates. Since the present
  sample deals with pairs and multiplets a fast destruction rate might be caused
  by the internal dynamical evolution in each cluster complex, caused by merger
  and/or other effects.

Approximately 55\% of cluster pairs and multiplets in Table~\ref{tab:interact}
present similar ages between their members indicating that they are coeval.
This suggests that most of the pairs and multiplets had a common origin,
possibly from the same molecular complex. Note that about 60\% of the pairs and
multiplets embedded in OB associations (H-A, Hodge 1985), as indicated in column
9 of Table~\ref{tab:catalog}, have comparable ages between their
members. This could be an explanation for the origin of cluster systems.

Considering triplet and multiplet members, we found that about 70\% of them are
younger than 100 Myr. These results suggest a possible binary cluster formation
scenario: clusters can be born in multiplet systems and coalesce by mergers and
tidal disruptions forming binary clusters in a timescale of $\approx$ 100
Myr. This time is in agreement with dynamical times required for an interacting
pair to merge into a single cluster (ODB98, de Oliveira et al. 2000).

The pairs with a bridge in the isophotal maps have comparable ages for their
components (Table~\ref{tab:interact}). As examples, the pairs NGC241/NGC242 and
B39/BS30 in Fig.~\ref{fig:atlas2} show a bridge linking their members which
could be interpreted as a sign of interaction (see the similarity with the
N-body simulation model in Fig. 11 of ODB98).  A typical timescale for the
bridge phenomenon is  $\approx$ 30 Myr, as deduced from a series of N-body simulations related to 
bridge formation and evolution (ODB98).

Another interesting isophotal feature is related to the cluster triplet
NGC220/NGC222/B23 which shows distortions for the small cluster in a direction
almost perpendicular to the line connecting itself to the large components NGC220 and
NGC222. This configuration and morphologies are compatible with a fast
hyperbolic encounter with small impact parameter (e.g. Fig. 12 of ODB98).

 Fig.~\ref{fig:distr_SMC} shows the angular distribution of pairs and multiplets
 together with SMC HI column density isophotes from Mathewson \& Ford (1984). It
 can be seen that  most of the objects are concentrated in the SMC main body,
 close to the higher concentration of HI, so it is not unexpected that in
 general they result young (Sect. 4). The objects appear to be distributed
 along an axis. Such distribution is present in the overall SMC cluster sample
 and there is growing evidence that it is related to a nearly edge-on disk
 containing the bulk of the young stellar population in the SMC (Bica {\it et
 al.} 1999, Westerlund 1990).

A nearly edge-on disk in a low internal reddening galaxy like the SMC would
imply an increase of projection effects as compared to a simulation such as that
carried out by Bhatia \& Hatzidimitriou (1988) for the nearly face-on LMC disk
where the physical pairs would be about 50\%. Consequently the fraction of
physical pairs in the SMC would be lower. The present approach including
morphological evidence of interaction is an attempt to constrain this aspect.
Indeed the fraction with isophotal distortions is only 25\% (Sect.
3.1). Projection effects can be responsible for the age spread in some
multiplets. For example the sextuplet (Table~\ref{tab:interact}) has component
ages 80 Myr (B72), 200 Myr (H86-143), 50 Myr (BS257), 3 Myr (SMC-N52A and
SMC-N52B) and 400 Myr (H86-148). Possibly only the 3 or 4 younger components
could be related to OB-Association H-A35, the remaining objects would be captures or
projection effects. This age spread is also present in  IC1611's quadruplet
and in some triplets. On the other hand the quintuplet in the star forming
complex NGC395 has all its members with the same age (3 Myr) thus  forming a
physical system.

In Fig.~\ref{fig:diam_SMC} we show the distribution of the diameter ratio between
members for all pairs in the sample. The diameter ratio is mostly in
the range $1 - 2$, with a peak at $1$ indicating that the majority of pair
members have  a comparable  size. This effect was also observed in the LMC (Bhatia et al. 1991).

Fig.~\ref{fig:separ_SMC} shows the distribution of the centre-to-centre angular
separation between pair members. The  separation range is $\approx$ $10
- 80$ arcsec ($\approx$3 - 22 pc) with a pronounced peak at $\approx$ $45$ arcsec (13 pc). A
similar peak was also observed by Bhatia  et al. (1991) and de Oliveira
(1996) who found a bimodal distribution for the LMC pairs with peaks at
$\approx$ 5 and 13 pc. The observed upper limit of the projected
centre-to-centre linear separation  $\approx$ 23 pc (80 arcsec) is comparable to 
Bhatia \& Hatzidimitriou's (1988) separation criterion for pairs in the LMC
(18.7 pc). The frequent  separation value around 13 pc may reflect a 
preferred survival distance for the systems, combined to projection effects.

\section{Conclusions}

We presented an isophotal atlas for 75 star cluster pairs and multiplets in the
Small Magellanic Cloud, comprising 176 objects.

 It was possible to derive ages from Colour-Magnitude Diagrams using the OGLE-II
  photometric survey for 91 objects. In addition we included in the analysis
  ages for 21 embedded objects in HII regions. The age distribution has a
  maximum in the youngest bin with a profile related to the cluster
  formation/destruction rates, in particular cluster multiple systems can have a
  fast destruction rate caused by their internal dynamical evolution. There is a
  second peak around 220 Myr which is probably related to the SMC/LMC last encounter.

 We find that 55\% of the pairs and multiplets
  in the sample are in general coeval, indicating that captures are a rare
  phenomenon. Most of the cluster multiplets occur in OB stellar
  associations and/or HII region complexes which indicates a common origin and
  suggests that multiplets coalesce into pairs or single clusters in a short
  time scale ($\approx$ 100 Myr).

 The majority of the cluster members have comparable sizes, with a diameter
ratio ranging mostly between $1 - 2$.  The projected separation distribution
between the members of a pair has a pronounced peak at $\approx$ 13 pc.  These
observational results are important constraints to theoretical models of star
cluster pair encounters and could be related with the formation process and
subsequent dynamical evolution of cluster systems.

The angular distribution of cluster pairs and multiplets shows that most of the
objects are located in the SMC main body. The overall SMC cluster sample
presents a similar distribution and there is evidence that it is related to a
nearly edge-on disk in the SMC. Considering this, it is expected an increase of
projection effects as compared to estimates for the LMC disk where physical
pairs would be about 50\% (Bhatia \& Hatzidimitriou 1988).

The atlas shows that around 25\% of the isophote maps present relevant
structures like bridges, common envelopes or detached distorted 
isophotes. N-body simulations have indicated that
these structures arise from interactions between the members of the cluster
systems. Indeed cluster pairs as NGC241/NGC242 and B39/BS30 show in their
isophotal maps bridges linking their members and have comparable ages. 

We conclude that multiplicity may have an important role in the early dynamical
evolution of star clusters in general, and signatures of that may survive in the
long term structure of large single clusters (de Oliveira  et al. 2000).

\begin{acknowledgements}

We thank the referee Dr. B. Westerlund for interesting remarks.
 We acknowledge support from the Brazilian institutions CNPq, CAPES and FINEP.

\end{acknowledgements}

\end{document}